\documentclass[sigconf]{acmart}

\usepackage{ctable}
\usepackage{multirow}
\usepackage{balance}
\usepackage{booktabs}
\usepackage{hyphsubst}
\usepackage{graphicx}
\usepackage{amsmath}
\usepackage{url}
\usepackage{breakurl}
\usepackage{hhline}
\usepackage[caption = false]{subfig}

\newcommand{\para}[1]{\vspace{2mm}\noindent\textbf{#1}}
\newcommand{\subpara}[1]{\textit{\textbf{#1}}}

\begin{document}

\title{The Impact of Semantic Context Cues on the User Acceptance of Tag Recommendations: An Online Study}

\author{Dominik Kowald*, Paul Seitlinger**, Tobias Ley**, Elisabeth Lex*}
\affiliation{
  \institution{*Know-Center Graz \& Graz University of Technology (Austria), **Tallinn University (Estonia)}
}
\email{dkowald@know-center.at, paul.seitlinger@tlu.ee, tley@tlu.ee, elisabeth.lex@tugraz.at}

\begin{abstract}
In this paper, we present the results of an online study with the aim to shed light on the impact that semantic context cues have on the user acceptance of tag recommendations. Therefore, we conducted a work-integrated social bookmarking scenario with 17 university employees in order to compare the user acceptance of a context-aware tag recommendation algorithm called \textit{3Layers} with the user acceptance of a simple popularity-based baseline. In this scenario, we validated and verified the hypothesis that semantic context cues have a higher impact on the user acceptance of tag recommendations in a \textit{collaborative tagging setting} than in an \textit{individual tagging setting}. With this paper, we contribute to the sparse line of research presenting online recommendation studies.

\para{Keywords.} Tag Recommendations; Online Evaluation; 3Layers
\end{abstract}

\copyrightyear{2018}
\acmYear{2018}
\setcopyright{iw3c2w3}
\acmConference[WWW '18 Companion]{The 2018 Web Conference Companion}{April 23--27, 2018}{Lyon, France}
\acmBooktitle{WWW '18 Companion: The 2018 Web Conference Companion, April 23--27, 2018, Lyon, France}
\acmPrice{}
\acmDOI{10.1145/3184558.3186899}
\acmISBN{978-1-4503-5640-4/18/04}

\fancyhead{}
\settopmatter{printacmref=false}

\maketitle

\section{Introduction}
Tag recommendation algorithms support users in finding descriptive tags for their bookmarked resources. Related research has proposed various ways to calculate tag recommendations (see e.g., \cite{cikm_3l}) and in our previous work \cite{Kowald2016a}, we have shown that the factors of tag usage frequency, recency and semantic context influence the accuracy of tag recommendations. While frequency and recency were found to be important in both individual tagging settings (i.e., narrow folksonomies) and collaborative tagging settings (i.e, broad folksonomies), the semantic context only increased the tag recommendation accuracy in the collaborative tagging setting.

Therefore, it is the aim of this paper to validate these offline evaluation results also in an online tag recommendation study, which enables us to test the following hypothesis: ``Semantic context cues have a higher impact on the user acceptance of tag recommendations in a \textit{collaborative tagging setting} than in an \textit{individual tagging setting}''. 
Thus, the contributions of our paper are two-fold: (i) we shed light on the impact that semantic context cues have on the user acceptance of tag recommendations, and (ii) with this paper, we contribute to the sparse line of research presenting results of online tag recommendation studies (see e.g., \cite{taglive}).


\section{Method}
\begin{table}[t!]
	\small
  \setlength{\tabcolsep}{5.0pt}	
  \centering
    \begin{tabular}{l|ccccc}
    \specialrule{.2em}{.1em}{.1em}
		Tagging setting		 & $|T|$				& $|Y|$			& $|B|$		& $|MostPop|$	& $|3Layers|$				\\\hline 
		Individual (narrow)				 &	119					&	191					&	53			&	29					&	24								\\\hline
		Collaborative	(broad)		 &	127					& 262			 		& 62			& 29					& 33								\\\hline
		Full dataset			 &	213					& 453			 		& 115			& 58					& 57								\\
		\specialrule{.2em}{.1em}{.1em}								
    \end{tabular}
    \caption{The data collected by the 17 participants during the four weeks of our online bookmarking study: $|T|$ is the number of distinct tags, $|Y|$ is the number of tag assignments, $|B|$ is the number of bookmarks, $|MostPop|$ is the number of bookmarks supported by the $MostPop$ algorithm and $|3Layers|$ is the number of bookmarks supported by the $3Layers$ algorithm.
\vspace{-7mm}}	
  \label{tab:datasets}
\end{table}

We constructed a work-integrated bookmarking scenario, in which 17 university employees from the areas of computer science and psychology have explored a given topic (i.e., ``designing workplaces that inspire people'') for a period of four weeks. Each participant was assigned the task to collect topic-relevant resources (i.e., three Web links or documents per week) in the DropBox-like tool \textit{KnowBrain}\footnote{\url{https://github.com/learning-layers/KnowBrain}}. Furthermore, they had to annotate these resources using a tagging interface.

In this tagging interface, the participants were not only asked to provide a set of tags but also to select from a predefined set of categories (i.e., ``Gamification \& Playfulness'', ``Inspiration sources \& techniques'', ``Collaboration technologies'', ``Personalization services'', ``Augmented reality'', ``Interior design'', ``Wellbeing \& health'', and ``Socializing''). These category selections were then used as semantic context cues for the calculation of tag recommendations.

To test our hypothesis, the participants were split into two groups, where the participants in the first group conducted the task individually and the participants of the second group conducted the task collaboratively. The collected data for both groups as well as for the full dataset is shown in Table \ref{tab:datasets}. For details on the study design, please see \cite{seitlinger2017hci}.

\begin{figure*}[t!]
   \centering
	 \captionsetup[subfigure]{justification=centering}
	 \subfloat[Individual tagging setting (narrow)]{ 
      \includegraphics[width=0.28\textwidth]{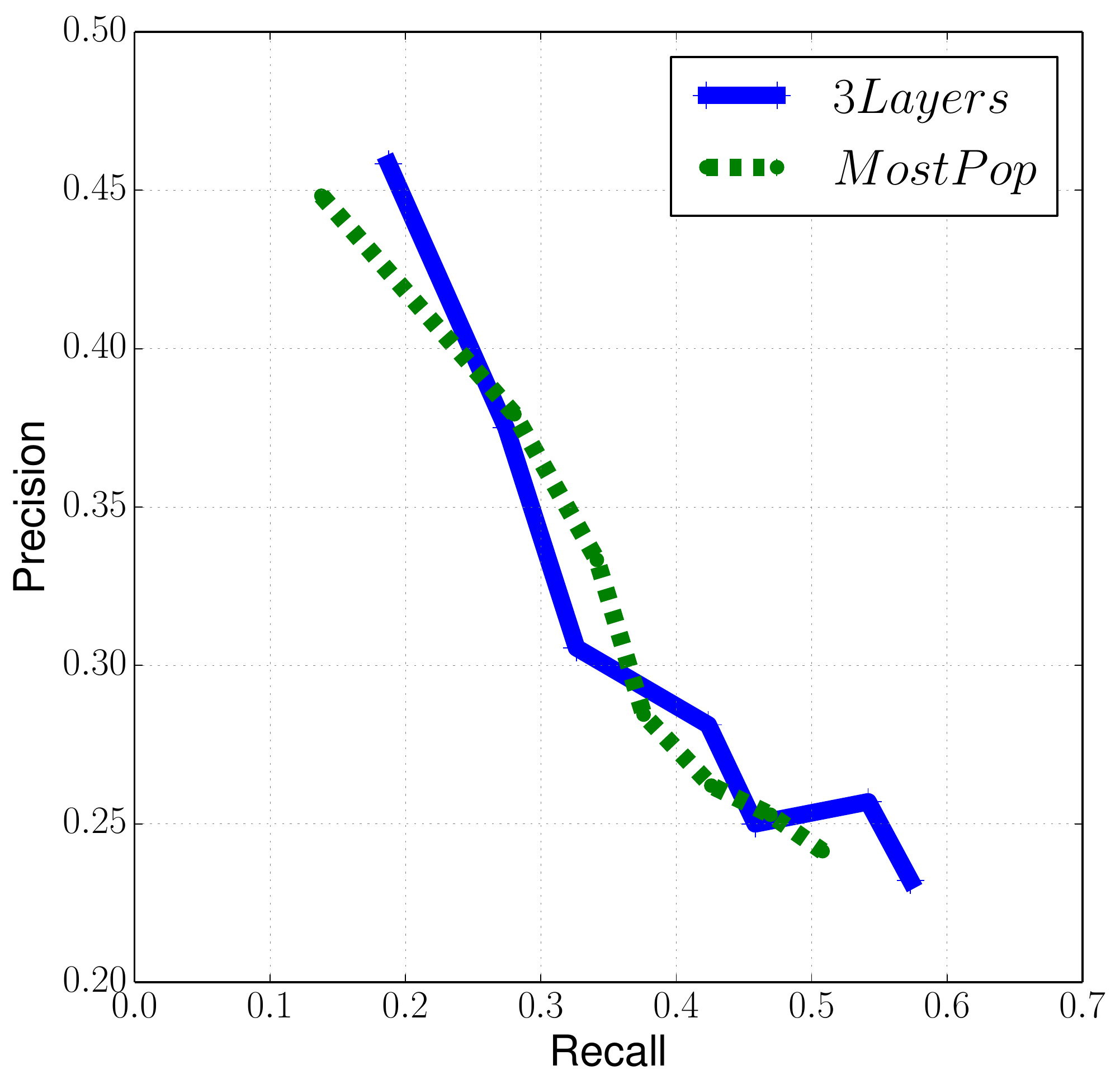} 
   }
	 \subfloat[Collaborative tagging setting (broad)]{ 
      \includegraphics[width=0.28\textwidth]{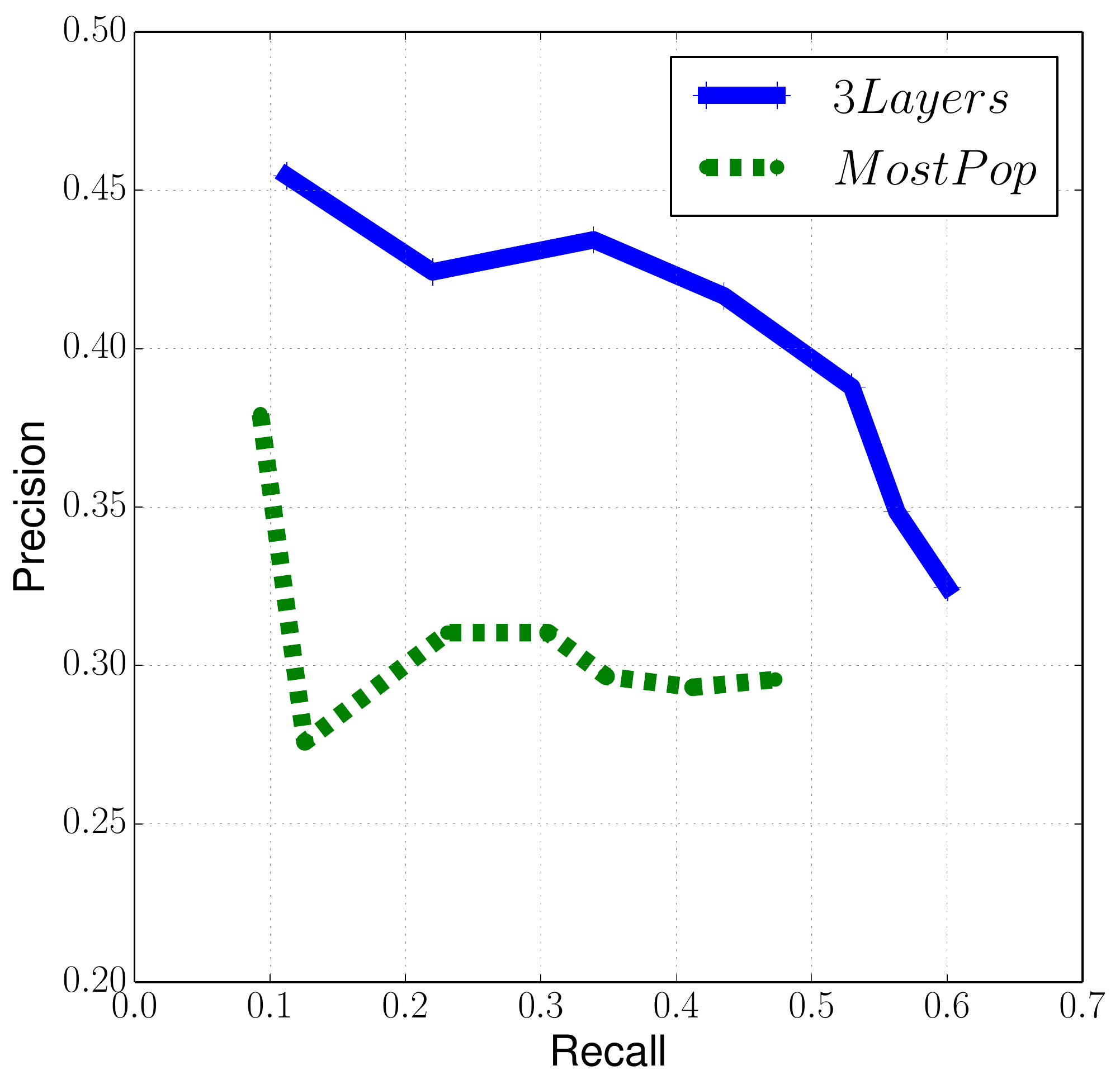} 
   }
	 \subfloat[Full dataset (narrow + broad)]{ 
      \includegraphics[width=0.28\textwidth]{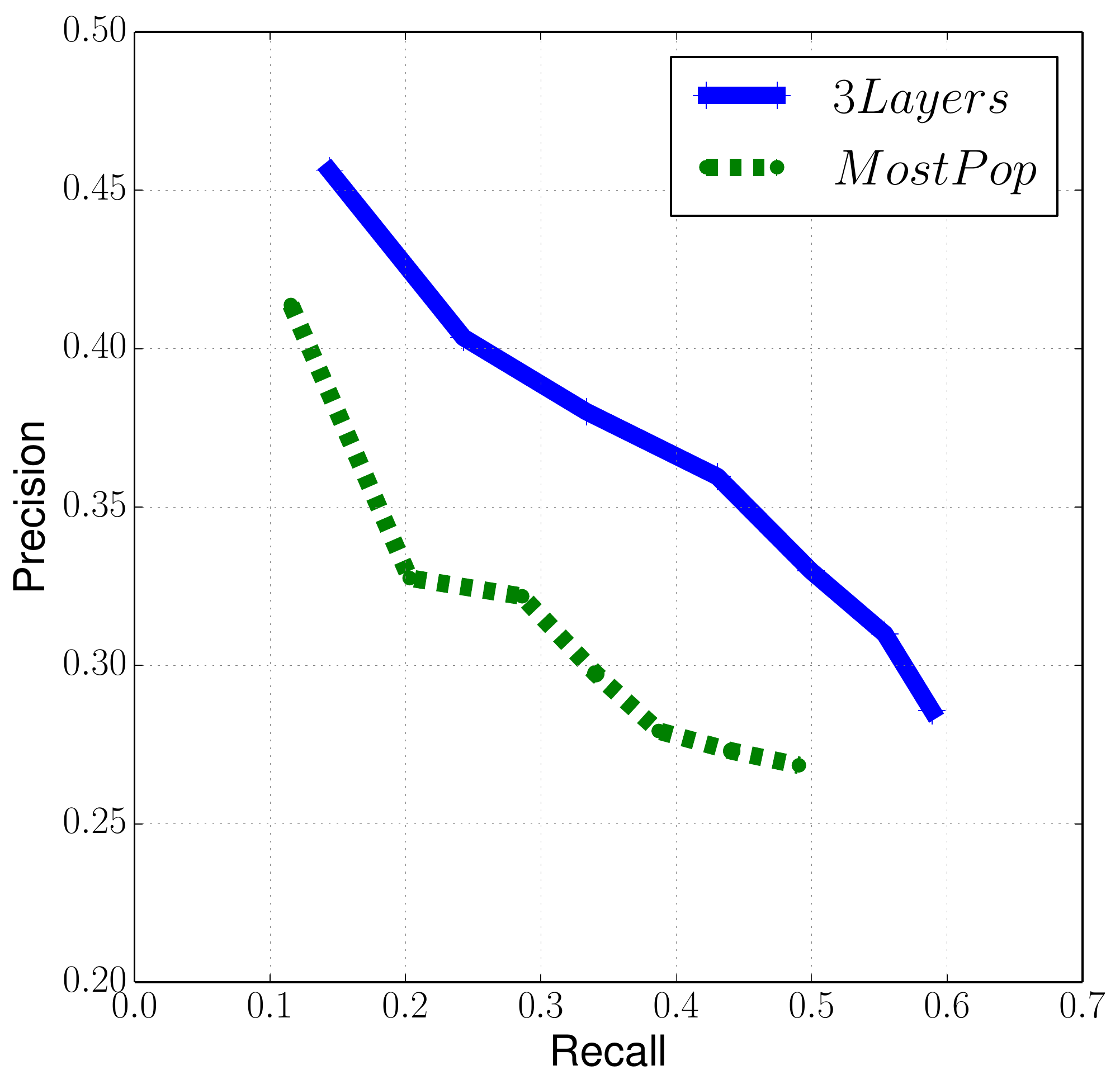} 
   }
   \caption{User acceptance rates for the two tagging settings (a and b), and the full dataset (c) measured by Precision and Recall. The results show that semantic context cues have a higher impact on these metrics in the collaborative setting than in the individual one.
\vspace{-3mm}}
	 \label{fig:results}
\end{figure*}

\subpara{Tag recommendations.} The tagging interface supported the annotation process by showing 7 tag recommendations to the users, which were either calculated via a baseline approach (i.e., $MostPop$) or an algorithm, which incorporates semantic context cues (i.e., $3Layers$). For each bookmark, the recommendation algorithm was chosen by the system at random.

\subpara{MostPop.} The $MostPop$ algorithm is a simple tag recommendation approach, which ranks the tags by their overall frequency. Thus, $MostPop$ recommends the 7 most frequently used tags.

\subpara{3Layers.}
3Layers builds on MINERVA2, which is a parameter-free formalization of how people make use of context cues (e.g., semantic categories) to (i) search memory for contextually similar episodes (e.g., of other bookmarks with similar category combinations), and (ii) access relevant items, such as tags that have frequently co-varied with these episodes (e.g., popular tags of similar bookmarks) \cite{hintzman1984minerva}. For implementing this context cue-dependent search process, we realized a simple ranking principle that weighs a tag's usage frequency by the extent to which it has co-occurred with categories that apply to the current resource (see \cite{cikm_3l}).

\subpara{User acceptance.} We measured the user acceptance by means of Precision and Recall for $k$ = 1 - 7 recommended tags. Hence, for each bookmark, we compared the set of recommended tags with the set of tags the user has actually used in the bookmark. Additionally, we report the F1-score for $k$ = 5 (i.e., $F1@5$) as also used in \cite{kowald2015evaluating}.

\section{Results}
In Figure \ref{fig:results}, we present the results of our online study related to the two tagging settings. Additionally, we present the results for the full dataset in order to relate our results to previous offline studies.

\subpara{Individual tagging setting.} In the individual setting (plot (a) of Figure \ref{fig:results}), we see overlapping Precision / Recall plot-lines and thus, cannot observe a significant difference between the user acceptance of $MostPop$ and $3Layers$. This is in line with the F1-score for $k$ = 5 since $F1@5$ = .309 for $3Layers$ and $F1@5$ = .302 for $MostPop$ ($p >$ .5)\footnote{According to a Wilcoxon rank-sum significance test.}. These results show that semantic context cues have only a small impact in an individual (i.e., narrow) tagging setting.

\subpara{Collaborative tagging setting.} In the collaborative setting (plot (b) of Figure \ref{fig:results}), we identify a significant advantage of $3Layers$ with $F1@5$ = .418 over $MostPop$ with $F1@5$ = .302 ($p <$ .05). We attribute this to the fact that in this collaborative setting, $3Layers$ is better able to exploit the additional information of the semantic context than in the individual tagging setting. These results mirror our previous work since in \cite{Kowald2016a}, we showed that the semantic context has a low influence in narrow folksonomies (i.e, in an individual setting) and a high influence in broad folksonomies (i.e., in a collaborative setting).

\subpara{Full dataset.} In the full dataset (plot (c) of Figure \ref{fig:results}), $3Layers$ with $F1@5$ = .372 significantly outperforms $MostPop$ with $F1@5$ = .302 ($p <$ .05). This validates our previous results presented in \cite{cikm_3l}, where we showed that $3Layers$ outperforms a context-unaware algorithm (i.e., Latent Dirichlet Allocation -- LDA) in an offline study setting, now also in an online study setting.

\section{Conclusion}
In this paper, we presented the results of an online study to identify the impact of semantic context cues on the user acceptance of tag recommendations. Based on our stated hypothesis, our two main findings showed (i) that the impact of semantic context cues greatly depends on the given tagging setting (i.e, individual vs. collaborative), and (ii) that in a collaborative setting, a tag recommendation algorithm, which incorporates semantic context cues, reaches a higher user acceptance than a solely frequency-based algorithm. These findings are also well in line with the results of previous offline tag recommendation evaluation studies \cite{cikm_3l,Kowald2016a}, which are now verified in an online setting. For future work, we plan to extend our study design by a third tag recommendation algorithm, which incorporates the temporal context of tag assignments (i.e., $BLL_{AC}$ from \cite{Kowald2016a}). We hypothesize, that this time-aware approach positively influences the user acceptance of recommendation in individual tagging settings.

\subpara{Acknowledgments.} This work is funded by the Know-Center, the EU projects Learning Layers (GA318209) and AFEL (GA687916), and the Austrian Science Fund (P25593-G22 and P27709-G22).


\begin{thebibliography}{6}


\ifx \showCODEN    \undefined \def \showCODEN     #1{\unskip}     \fi
\ifx \showDOI      \undefined \def \showDOI       #1{#1}\fi
\ifx \showISBNx    \undefined \def \showISBNx     #1{\unskip}     \fi
\ifx \showISBNxiii \undefined \def \showISBNxiii  #1{\unskip}     \fi
\ifx \showISSN     \undefined \def \showISSN      #1{\unskip}     \fi
\ifx \showLCCN     \undefined \def \showLCCN      #1{\unskip}     \fi
\ifx \shownote     \undefined \def \shownote      #1{#1}          \fi
\ifx \showarticletitle \undefined \def \showarticletitle #1{#1}   \fi
\ifx \showURL      \undefined \def \showURL       {\relax}        \fi
\providecommand\bibfield[2]{#2}
\providecommand\bibinfo[2]{#2}
\providecommand\natexlab[1]{#1}
\providecommand\showeprint[2][]{arXiv:#2}

\bibitem[\protect\citeauthoryear{Hintzman}{Hintzman}{1984}]%
        {hintzman1984minerva}
\bibfield{author}{\bibinfo{person}{Douglas~L Hintzman}.}
  \bibinfo{year}{1984}\natexlab{}.
\newblock \showarticletitle{MINERVA 2: A simulation model of human memory}.
\newblock \bibinfo{journal}{\emph{Behavior Research Methods, Instruments, \&
  Computers}} \bibinfo{volume}{16}, \bibinfo{number}{2} (\bibinfo{year}{1984}),
  \bibinfo{pages}{96--101}.
\newblock


\bibitem[\protect\citeauthoryear{J\"{a}schke, Eisterlehner, Hotho, and
  Stumme}{J\"{a}schke et~al\mbox{.}}{2009}]%
        {taglive}
\bibfield{author}{\bibinfo{person}{Robert J\"{a}schke}, \bibinfo{person}{Folke
  Eisterlehner}, \bibinfo{person}{Andreas Hotho}, {and} \bibinfo{person}{Gerd
  Stumme}.} \bibinfo{year}{2009}\natexlab{}.
\newblock \showarticletitle{Testing and Evaluating Tag Recommenders in a Live
  System}. In \bibinfo{booktitle}{\emph{Proc. of RecSys'09}}.
  \bibinfo{publisher}{ACM}, \bibinfo{address}{New York, NY, USA},
  \bibinfo{pages}{369--372}.
\newblock
\showISBNx{978-1-60558-435-5}


\bibitem[\protect\citeauthoryear{Kowald and Lex}{Kowald and Lex}{2015}]%
        {kowald2015evaluating}
\bibfield{author}{\bibinfo{person}{Dominik Kowald} {and}
  \bibinfo{person}{Elisabeth Lex}.} \bibinfo{year}{2015}\natexlab{}.
\newblock \showarticletitle{Evaluating tag recommender algorithms in real-world
  folksonomies: A comparative study}. In \bibinfo{booktitle}{\emph{Proc. of
  RecSys'15}}. ACM, \bibinfo{pages}{265--268}.
\newblock


\bibitem[\protect\citeauthoryear{Kowald and Lex}{Kowald and Lex}{2016}]%
        {Kowald2016a}
\bibfield{author}{\bibinfo{person}{Dominik Kowald} {and}
  \bibinfo{person}{Elisabeth Lex}.} \bibinfo{year}{2016}\natexlab{}.
\newblock \showarticletitle{The Influence of Frequency, Recency and Semantic
  Context on the Reuse of Tags in Social Tagging Systems}. In
  \bibinfo{booktitle}{\emph{Proc. of HT'16}}. ACM, \bibinfo{pages}{237--242}.
\newblock


\bibitem[\protect\citeauthoryear{Seitlinger, Kowald, Trattner, and
  Ley}{Seitlinger et~al\mbox{.}}{2013}]%
        {cikm_3l}
\bibfield{author}{\bibinfo{person}{Paul Seitlinger}, \bibinfo{person}{Dominik
  Kowald}, \bibinfo{person}{Christoph Trattner}, {and} \bibinfo{person}{Tobias
  Ley}.} \bibinfo{year}{2013}\natexlab{}.
\newblock \showarticletitle{Recommending tags with a model of human
  categorization}. In \bibinfo{booktitle}{\emph{Proc. of CIKM'13}}. ACM,
  \bibinfo{pages}{2381--2386}.
\newblock


\bibitem[\protect\citeauthoryear{Seitlinger, Ley, Kowald, Theiler,
  Hasani-Mavriqi, Dennerlein, Lex, and Albert}{Seitlinger
  et~al\mbox{.}}{2017}]%
        {seitlinger2017hci}
\bibfield{author}{\bibinfo{person}{Paul Seitlinger}, \bibinfo{person}{Tobias
  Ley}, \bibinfo{person}{Dominik Kowald}, \bibinfo{person}{Dieter Theiler},
  \bibinfo{person}{Ilire Hasani-Mavriqi}, \bibinfo{person}{Sebastian
  Dennerlein}, \bibinfo{person}{Elisabeth Lex}, {and} \bibinfo{person}{Dietrich
  Albert}.} \bibinfo{year}{2017}\natexlab{}.
\newblock \showarticletitle{Balancing the Fluency-Consistency Tradeoff in
  Collaborative Information Search with a Recommender Approach}.
\newblock \bibinfo{journal}{\emph{International Journal of Human–Computer
  Interaction}} (\bibinfo{year}{2017}), \bibinfo{pages}{1--19}.
\newblock


\end{thebibliography}

\end{document}